\documentclass[twocolumn, aps, prb]{revtex4}
\usepackage{graphicx}
\usepackage{amssymb}

\def\Vec#1{{\bf #1}}
\def\GVec#1{\mbox{\boldmath $#1$}}

\def\H{{\mathcal H}}
\def\vare{\varepsilon}
\def\av#1{\langle #1 \rangle}

\begin{document}

\title{Transport in Bilayer Graphene: Calculations within a self-consistent Born approximation}
\author{Mikito Koshino and Tsuneya Ando}
\affiliation{
Department of Physics, Tokyo Institute of Technology
2-12-1 Ookayama, Meguro-ku, Tokyo 152-8551, Japan}
\date{\today}

\begin{abstract}
The transport properties of a bilayer graphene are studied theoretically within a self-consistent Born approximation.
The electronic spectrum is composed of
$k$-linear dispersion in the low-energy region
and $k$-square dispersion as in an ordinary two-dimensional metal at high
energy, leading to a crossover between different behaviors in the
conductivity on changing the Fermi energy or disorder strengths. We
find that the conductivity approaches
$2e^2/\pi^2\hbar$ per spin in the strong-disorder regime,
independently of the short- or long-range disorder.
\end{abstract}

\maketitle

\section{Introduction}

Recently there was an experimental development in fabrication of 
atomically thin graphene, or single-layer graphite, that enables us to 
access its exotic electronic properties.\cite{Novo04,Berg,Zhan05-1} The 
magnetotransport was measured and the integer quantum Hall effect was 
observed.\cite{Novo05,Zhan05-2,Novoselov_et_al_2006a} In the experiments, 
a multilayer that contains a few graphene sheets is also 
available.\cite{Novo05,Novoselov_et_al_2006a} The electronic structure 
of a bilayer graphene was studied theoretically and the spectrum was 
found to be essentially different from that of a monolayer.\cite{McCa} 
The purpose of this paper is to study transport properties of the 
bilayer graphene.

A graphite monolayer has a $k$-linear, massless Dirac-like spectrum and 
has long attracted theoretical interests as a ``relativistic'' problem 
in condensed matter physics, where $k$ is the absolute value of the wave 
vector. Theoretical studies of transport in such an exotic electronic 
structure have been given by several authors, where the conductivity 
with or without magnetic field,\cite{Shon, Pere} the Hall 
effect,\cite{Zhen, Gusy} quantum corrections to the 
conductivity,\cite{Suzuura_and_Ando_Antilocalization} and the dynamical 
transport\cite{Suzu} are investigated. The results show that the 
conductivity exhibits various singular behaviors in the vicinity of zero 
energy.
\cite{Shon,Zhen,Suzu,Ando_and_Suzuura_2003a,
Ando_2003-2004:_Quantum_anomalies,Frad}

In bilayer graphene the energy dispersion includes both $k$-linear 
and $k$-square terms. In this paper, we calculate the diagonal 
conductivity in the absence of a magnetic field within a self-consistent 
Born approximation (SCBA). A similar SCBA analysis has been applied for 
monolayer graphene.\cite{Shon,Zhen,Suzu} We shall find that the 
coexistence of $k$-linear and $k$-square dispersions leads to a 
crossover between transport properties 
similar to those of a monolayer graphene and 
to those of 
an ordinary two-dimensional metal as the Fermi energy is changed. 
The conductivity becomes nearly universal, $2e^2/\pi^2\hbar$ per spin, 
in the case of large disorder. The analysis is made for two different 
kinds of scatterers, short range and long range, where the former 
represents on-site random energies distributed on the carbon atoms and 
the latter a slowly varying random potential 
of the range much longer than the lattice 
constant but shorter than the typical electron wavelength.

In Sec.\ II, the effective Hamiltonian and the resulting energy spectrum 
in a bilayer graphene are discussed, and model scatterers are introduced. 
The self-consistent Born approximation is briefly described in Sec.\ III. 
 Explicit results are discussed for short-range scatterers in Sec.\ IV 
and for long-range scatterers in Sec.\ V.  A discussion and a brief 
summary are given in Sec.\ VI.

\begin{figure}
\begin{center}
\begin{tabular}{cc}
 \leavevmode\includegraphics[width=80mm]{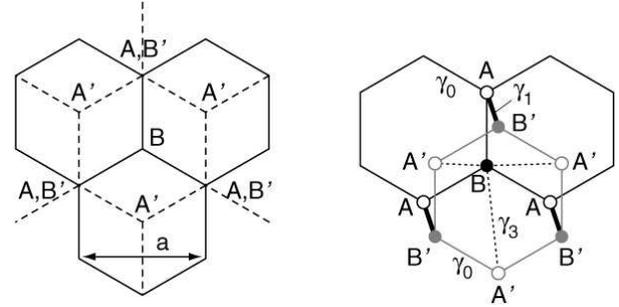}
\end{tabular}
\end{center}
\caption{(Left) Top view of the atomic structure in a bilayer graphene.
Solid and dashed lines represent the top ($A$ and $B$ sites) and bottom
 layers ($A'$ and $B'$ sites), respectively.
(Right) Definition of the hopping parameters in a tight-binding model, $\gamma_0$ between nearest-neighbor sites in each layer, $\gamma_1$ between $A$ and $B'$, and $\gamma_3$ between $B$ and $A'$.
}
\label{fig_bilayer}
\end{figure}

\section{Effective Hamiltonian and Energy Spectrum}

The bilayer graphene is composed of a pair of hexagonal networks of 
carbon atoms, which include $A$ and $B$ atoms on the top layer and $A'$ 
and $B'$ on the bottom, as schematically shown in Fig.\ \ref{fig_bilayer}. 
The two layers are arranged in AB stacking, where $A$ atoms are located 
above $B'$ atoms, and $B$ or $A'$ atoms are above or below the center of 
hexagons in the other layers. The unit cell contains four atoms $A$, $B$, 
$A'$, and $B'$, and the Brillouin zone becomes identical with that of 
the monolayer graphene. We model the system by a tight-binding 
Hamiltonian based on the Slonczewski-Weiss-McClure graphite 
model.\cite{McCl,Slon} We include three parameters $\gamma_0$, 
$\gamma_1$, and $\gamma_3$, where $\gamma_0$ represents the intralayer 
coupling $A\leftrightarrow B$ or $A'\leftrightarrow B'$, and $\gamma_1$ 
and $\gamma_3$ the interlayer coupling $A\leftrightarrow B'$ and 
$B\leftrightarrow A'$, respectively. The coupling parameters are 
estimated as $\gamma_0 \approx 3.16$ eV,\cite{Toy} $\gamma_1 \approx 
0.39$ eV,\cite{Misu} and $\gamma_3 \approx 0.315$ eV.\cite{Doez}

We can show that the low energy spectrum is given by the states around 
the $K$ and $K'$ points in the Brillouin zone. Neighboring $A$ and $B'$ 
sites are coupled by $\gamma_1$ to create the bonding and antibonding 
states away from the Fermi level, and the low energy states are given by 
the remaining $A'$ and $B$ sites.\cite{McCa} The effective Hamiltonian reads,
\begin{eqnarray}
\H_K &=& \frac{\hbar^2}{2m^*} \left( \begin{array}{cc}  0 & k_-^2 \\ k_+^2 & 0 \end{array} \right) - \frac{\hbar^2 k_0}{2m^*} \left( \begin{array}{cc}  0 & k_+ \\ k_- & 0 \end{array} \right), 
\\
\H_{K'} &=& \frac{\hbar^2}{2m^*} \left( \begin{array}{cc} 0 & k_+^2 \\ k_-^2 & 0 \end{array} \right) + \frac{\hbar^2 k_0}{2m^*} \left( \begin{array}{cc}  0 &  k_- \\  k_+ & 0 \end{array} \right),
\label{eq_Heff}
\end{eqnarray}
where $k_\pm = k_x \pm i k_y$ with $\Vec{k}$ being the wave vector 
measured from the $K$ or $K'$ points, and the effective mass $m^*$ and the 
wave number $k_0$ defined by
\begin{equation}
\frac{\hbar^2}{2m^*} = \frac{(\sqrt 3 a\gamma_0/2)^2}{\gamma_1},
\end{equation}
and
\begin{equation}
k_0 = \frac{2}{\sqrt 3 a}\frac{\gamma_3\gamma_1}{\gamma_0^2},
\label{eq_k0}
\end{equation}
with the lattice constant $a=0.246$ nm. The $k$-linear term in the 
Hamiltonian (\ref{eq_Heff}) describes the direct hopping between $A'$ 
and $B$ sites, and the $k$-square term the second-order process between 
$A'$ and $B$ via $A$-$B'$ dimers. A typical energy where the $k$-square 
and $k$-linear terms become comparable can be defined by
\begin{equation}
\vare_0 = \frac{\hbar^2k_0^2}{2m^*}  =
 \left(\frac{\gamma_3}{\gamma_0}\right)^2\gamma_1 .
\end{equation}

The eigenenergy of (\ref{eq_Heff}) becomes
\begin{eqnarray}
\vare_{j{\bf k}s} = \frac{\hbar^2}{2m^*} sk \sqrt{k^2 \mp 2k_0k \cos 3\varphi + k_0^2},
\label{eq_E}
\end{eqnarray}
where the upper sign corresponds to $j=K$ and the lower to $K'$, $s=\pm 1$, $k=\sqrt{k_x^2+k_y^2}$, and $\varphi = \arg(k_+)$ with $\arg (z)$ being the argument $\varphi$ in $z = |z|e^{i\varphi}$.
The eigenvectors corresponding to (\ref{eq_E}) are 
\begin{eqnarray}
\GVec{\phi}^{j{\bf k}s} = \left( \begin{array}{c} \phi_{A'}^{j{\bf k}s} \\ \noalign{\vspace{0.10cm}} \phi_B^{j{\bf k}s} \end{array} \right) = \frac{1}{\sqrt{2}} \left( \begin{array}{c}  e^{i\theta_{j{\bf k}}}\\ s \end{array} \right),
\end{eqnarray}
with
\begin{eqnarray}
\theta_{K{\bf k}} &=& \arg (-k_0 e^{i\varphi} + k e^{-2i\varphi}) ,
\\
\theta_{K'{\bf k}} &=& \arg (k_0 e^{-i\varphi} +  k e^{2i\varphi}).
\end{eqnarray}

\begin{figure}
\begin{center}
\begin{tabular}{cc}
 \leavevmode\includegraphics[width=60mm]{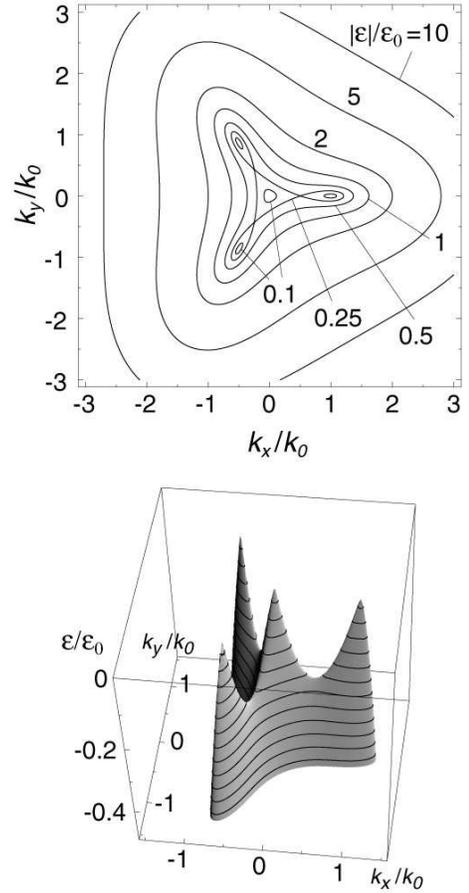}
\end{tabular}
\end{center}
\caption{Equienergy lines (top) and the 
three-dimensional plot (bottom) of the energy
dispersion of the bilayer graphene around the $K$ point.
In the latter only the lower half $(\vare < 0)$ is shown.}
\label{fig_disp}
\end{figure}

Figure \ref{fig_disp} shows the energy dispersion (\ref{eq_E}) for $j = 
K$. In the high-energy region $|\vare| > (1/4)\vare_0$, we have a single 
trigonally warped Fermi line, which becomes closer to a circle on going 
to the higher energy as the $k$-square term dominates in the Hamiltonian. 
The equienergy line becomes convex for $\vare \gtrsim 10.8\vare_0$. In 
the low-energy region $|\vare| < (1/4)\vare_0$, the Fermi line splits 
into four separate pockets for each of $K$ and $K'$, one center part 
and three satellite parts that are located trigonally. In the vicinity 
of zero energy $|\vare| \ll \vare_0$, the dispersion becomes linear in 
$k$ space with respect to the four Fermi points, where the center pockets 
can be approximated as a circle with radius $k = |\vare/\vare_0|k_0$ and 
the three satellites as ellipses with the longer and shorter radii 
$|\vare/\vare_0|k_0$ and $|\vare/\vare_0|k_0/3$. While splitting of 
the four 
Fermi lines occurs only at very low energies, the trigonal warping 
extends to much higher energy as seen in Fig.\ \ref{fig_disp}.

The velocity operator
$v_x = (\partial {\mathcal H}/\partial k_x)/\hbar$
has nonzero matrix elements 
between the states on the identical $k$ points $(j,\Vec{k})$
written as
\begin{equation}
(v_{j{\bf k}})_{ss'} \equiv \langle j {\bf k}s' | v_x | j {\bf k}s \rangle 
= sw_{j{\bf k}} + s'w^*_{j{\bf k}} ,
\label{eq_v}
\end{equation}
with
\begin{eqnarray}
 w_{K{\bf k}} &=& \frac{\hbar}{4m^*} 
(-k_0 + 2 k e^{-i\varphi})e^{-i\theta_{K{\bf k}}},
\\
 w_{K'{\bf k}} &=& \frac{\hbar}{4m^*} 
(k_0 + 2 k e^{i\varphi})e^{-i\theta_{K'{\bf k}}}.
\end{eqnarray}
In the high-energy region $|\vare| > (1/4)\vare_0$, the mean square velocity averaged on the contour 
$\vare = |\vare_{j\Vec{k}s}|$ is given by
\begin{equation}
\av{|(v_{j\Vec{k}})_{ss'}|^2}_{\vare} = 
\frac{|\vare|}{m^*} + \frac{\vare_0}{2m^*}(1+ss'), 
\label{eq_vsq_high}
\end{equation}
where the first and second terms come from the $k$-square and $k$-linear terms in the dispersion, respectively.
In the vicinity of zero energy $|\vare| \ll \vare_0$, we have
\begin{equation}
\label{eq_vsq_low}
\av{|(v_{j\Vec{k}})_{ss'}|^2}_{\vare} = \frac{3\vare_0}{4m^*}. 
\end{equation}

The density of states per spin is defined by
\begin{equation}
\rho_0(\varepsilon) = {1\over\Omega} \sum_{j{\bf k}s} \delta(\varepsilon-\varepsilon_{j{\bf k}s}) ,
\end{equation}
where $\Omega$ is the area of the system.
In the high energy region $|\vare| > (1/4)\vare_0$, this becomes
\begin{equation}
\rho_0(\varepsilon) \approx \frac{m^*}{\pi\hbar^2} \equiv \rho_{\infty} ,
\label{eq_rho_high}
\end{equation}
with terms of the order of $O(\vare_0^2/\vare^2)$ neglected.
In the vicinity of zero energy $|\vare| \ll \vare_0$, on the other hand,
\begin{equation}
\rho_0(\varepsilon) \approx \frac{4m^*}{\pi\hbar^2} \frac{|\vare|}{\vare_0} .
\label{eq_rho_low}
\end{equation}
The density of states diverges logarithmically at $|\varepsilon|/\varepsilon_0=1/4$ due to the presence of saddle points in the dispersion.

For the parameters mentioned above, $m^*/m_0=0.033$ with $m_0$ being the
free-electron mass, $k_0(2\pi/a)^{-1}=2.2\times10^{-3}$, and
$\varepsilon_0\!\approx\!3.9$ meV.  The electron concentration
corresponding to $(1/4)\varepsilon_0$ is $n_s=1.7\times10^{10}$ cm$^{-2}$, that
to $\varepsilon_0$ is $1.0\times10^{11}$ cm$^{-2}$, and that to
$10.8\varepsilon_0$ is $1.1\times10^{12}$ cm$^{-2}$.  As the typical electron
concentration in the present system is $10^{12}$ cm$^{-2}$ ($\varepsilon/\varepsilon_0\!\sim\!10$) or larger,\cite{Novoselov_et_al_2006a} the trigonal warping is appreciable, but it is extremely hard to realize the
situation where four Fermi lines are well split from each other.

The Hamiltonian (\ref{eq_Heff}) is formally equivalent to that of the 
monolayer system, where the $k$-square term comes from a higher-order 
term in the $\Vec{k}\cdot\Vec{p}$ approximation.\cite{Ajik} However, the 
parameter $\vare_0$ for the monolayer becomes the order of $\gamma_0$ 
and thus is larger than in the bilayer by the order of 1000, and the 
$k$-square term gives only a small perturbation. In the bilayer, on the 
contrary, the $k$-square term 
becomes dominant in the dispersion at small energy 
$\varepsilon_0$. Note also that the Hamiltonian (\ref{eq_Heff}) becomes 
invalid when the energy becomes as high as the antibonding states of 
the $A$-$B'$ dimers. The deviation from (\ref{eq_Heff}) seems to become 
appreciable around$\varepsilon\sim\gamma_1/4$ ($\sim0.1$ eV 
corresponding to$n_s\sim2.7\times10^{12}$ cm$^{-2}$).\cite{McCa} The 
dispersion of the bilayer graphene is much closer to that of 
three-dimensional graphite, where we see a similar trigonal structure in 
the $k_x-k_y$ plane with a fixed wave number $k_z$ along the stacking 
direction.\cite{McCl,Char}

In terms of the eigenvector ${\Vec \phi}^{j{\bf k}s}$ the amplitudes of
the atomic orbitals at sites ${\bf R}^{A'}$ and ${\bf R}^B$ are
given by
\begin{eqnarray}
 \psi_{A'}(\Vec{R}^{A'}) &=&
  {1\over\sqrt N} \phi_{A'}^{j{\bf k}s} 
\exp[i(\Vec{K}_j + \Vec{k})\cdot\Vec{R}^{A'}] ,
\\
\psi_{B}(\Vec{R}^{B}) &=& {\omega_j \over \sqrt N} 
\phi_{B}^{j{\bf k}s} \exp[i(\Vec{K}_j + \Vec{k})\cdot\Vec{R}^{B}] ,
\end{eqnarray}
for $j=K$ and $K'$, where $\omega_K = \omega^{-1}$, $\omega_{K'} = \omega$ with $\omega = \exp(2\pi i/3)$, and $N$ is the number of unit cells in the system.

The dominant scatterers in the present system are not well known.  In
the following, we shall consider two kinds of model scatterers,
short-range scatterers localized only on $B$ sites or on $A'$ sites and
long-range scatterers, the potential of which spreads over a certain
length scale larger than the lattice constant $a$.  In each case we
assume that the disorder strength is weaker than the dimer coupling
$\gamma_1$, so that we need to consider only the potential on $A'$ and
$B$ sites.  The effective Hamiltonian for the disorder potential can be
derived similarly as in monolayer graphene, if we identify the $A'$ and
$B$ sites in the bilayer with $A$ and $B$ in the
monolayer.\cite{Ando_and_Nakanishi_1998a,Ando_2005a}

The matrix elements for the short-ranged potential are written as
\begin{eqnarray}
&& \hspace{-0.50cm} \langle j'\Vec{k'}s'| U^{A'}_i |j\Vec{k}s\rangle \nonumber\\
&& = \frac{u_i^A}{\Omega}
e^{i(\Vec{K}_j + \Vec{k}
- \Vec{K}_{j'} - \Vec{k}')\cdot\Vec{R}^{A'}_i}
(\phi^{j'\Vec{k}'s'}_{A'})^*\phi^{j\Vec{k}s}_{A'} ,
\\
&& \hspace{-0.50cm} \langle j'\Vec{k'}s'| U^{B}_i |j\Vec{k}s\rangle \nonumber\\
&& = \frac{u_i^B}{\Omega}
e^{i(\Vec{K}_j + \Vec{k}
- \Vec{K}_{j'} - \Vec{k}')\cdot\Vec{R}^{B}_i}
(\omega_{j'}\phi^{j'\Vec{k}'s'}_{B})^*\omega_j\phi^{j\Vec{k}s}_{B} ,
\qquad
\end{eqnarray}
for scatterers at $\Vec{R}^{A'}_i$ and $\Vec{R}^{B}_i$, respectively,
where $u_i^A$ and $u_i^B$ are the integrated intensities of the
potential, i.e., the coefficient of the $\delta$ potential.  For long
range disorder, we assume that the potential range is much larger than
the lattice constant $a$ but smaller than the typical wavelength
$2\pi/k$ with $k$ being the wave number from the $K$ or $K'$ point.
Then we can neglect the matrix elements for intervalley scatterings
between $K$ and $K'$, while those for intravalley can be written as
\begin{equation}
\langle j'\Vec{k'}s'| U_i |j\Vec{k}s\rangle
= 
\delta_{jj'}
\frac{u_i}{\Omega}
e^{i(\Vec{k} - \Vec{k}')\cdot\Vec{R}_i}
(\GVec{\phi}^{j\Vec{k}'s'})^\dagger \GVec{\phi}^{j\Vec{k}s},
\end{equation}
where $u_i$ is the integrated intensity of the potential.

We have another possibility for the long-range disorder potential,
where each of the scatterers is effective only in one layer.
This is modeled by a long-range potential that has amplitude
only over either $A'$ or $B$ sites.
The situation then becomes almost equivalent to the short-range case, 
the only difference being 
that $K$ and $K'$ are decoupled, reducing the self-energy by a factor of
2.

\section{Self-Consistent Born Approximation}

In the self-consistent Born approximation, the self-energy of the
disorder-averaged Green's function $\langle G_{\alpha,\alpha'}\rangle$
is given by\cite{Shon}
\begin{equation}
\Sigma_{\alpha,\alpha'} (\vare) = 
\sum_{\alpha_1, \alpha_1'}
\langle 
U_{\alpha,\alpha_1}U_{\alpha'_1,\alpha'}
\rangle
\langle 
G_{\alpha_1,\alpha'_1}(\vare)
\rangle,
\end{equation}
with $\alpha= (j\Vec{k}s)$, where $\av{\ \ }$ 
represents the average over the impurity configurations.
In the present system the conductivity becomes isotropic in spite of the presence of strong trigonal warping.
It can be calculated by the Kubo formula,
\begin{equation}
 \sigma(\vare) = \frac{\hbar e^2}{2\pi\Omega} {\rm Re} \,\, {\rm Tr}
\left[
\av{v_x G^R v_x G^A} - \av{v_x G^R v_x G^R}
\right] ,
\end{equation}
where $G^R=(\vare - \H + i0)^{-1}$ and $G^A=(\vare - \H - i0)^{-1}$ are the retarded and the advanced Green's functions, respectively, with $\H$ being the Hamiltonian including the disorder potential.
This can be rewritten as
\begin{equation}
 \sigma(\vare) = \frac{\hbar e^2}{2\pi\Omega} {\rm Re} \,\, {\rm Tr}
\left[v_x \av{G^R} \tilde{v}_x^{RA} \av{G^A} 
-v_x \av{G^R} \tilde{v}_x^{RR} \av{G^R} 
\right],
\end{equation}
with $\tilde{v}_x^{RA} =  \tilde{v}_x(\vare+i0,\vare-i0)$
and $\tilde{v}_x^{RR} =  \tilde{v}_x(\vare+i0,\vare+i0)$ satisfying
\begin{equation}
 \tilde{v}_x(\vare,\vare') =  v_x + \av{UG(\vare)\tilde{v}_xG(\vare')U}.
\label{eq_vertex}
\end{equation}
In the SCBA, $\tilde v_x$ should be calculated in the ladder approximation.
In the above and hereafter we omit the summation over the spin degeneracy,
so the actual conductivity should be multiplied by a factor of 2.

In the case of the present model scatterers, the self-energy and
therefore the averaged Green's function become diagonal with respect to
the wave number and the band index.  Further, the self-energy is
independent of the wave number and the band index, and thus is determined
by the energy alone.  As a result we have
\begin{eqnarray}
&& \av{G_{\alpha,\alpha '}(\vare)} = \delta_{\alpha,\alpha '} G_\alpha(\vare) ,
\\
&& G_\alpha(\vare) = G(\vare,\varepsilon_\alpha) \equiv \frac{1}{\vare - \Sigma(\vare) - \vare_\alpha} ,
\end{eqnarray}
where $\Sigma(\varepsilon)$ is the self-energy.

\section{Short-range scatterers}

For short-range scatterers, we assume that they are equally distributed
to $A$ and $B$ sites with density $n_i^A = n_i^B = n_i/2$ and the
identical mean square amplitude $\av{(u_i^A)^2}=\av{(u_i^B)^2}=u^2$.
Then, the self-energy is given by
\begin{eqnarray}
\Sigma(\vare) = \frac{n_i u^2}{4\Omega} \sum_{\alpha} G_\alpha (\vare).
\label{eq_self0}
\end{eqnarray}
By substituting the summation over $j$ and $\Vec{k}$ 
with the integration in energy $\varepsilon'=|\varepsilon_{j{\bf k}s}|$,
we can rewrite this as
\begin{eqnarray}
 \Sigma(\vare) = \frac{n_i u^2}{4} \int_0^\infty d\vare ' \rho_0(\vare ')
\sum_s  G(\vare, s\vare') .
\label{eq_self}
\end{eqnarray}
The density of states is given by
\begin{equation}
 \rho(\vare) = -\frac{1}{\pi\Omega}\sum_\alpha {\rm Im} G_\alpha(\vare+i0)
= -\frac{4}{\pi n_iu^2}{\rm Im}\Sigma(\vare+i0).
\label{eq_dos}
\end{equation}

For the conductivity,
we can show that the vertex correction vanishes in the short-range
scatterers, or $\tilde{v_x} = v_x$, and we obtain
\begin{eqnarray}
 \sigma(\vare) &=& 
\frac{\hbar e^2}{2\pi\Omega} {\rm Re}\sum_{j\Vec{k}ss'}
|(v_{j\Vec{k}})_{ss'}|^2 
 (G^R_{j\Vec{k}s}G^A_{j\Vec{k}s'}-G^R_{j\Vec{k}s}G^R_{j\Vec{k}s'})
\nonumber\\
&& =  \frac{\hbar e^2}{2\pi}
\int_0^\infty d\vare' \rho_0(\vare')\sum_{ss'}
\av{|(v_{j\Vec{k}})_{ss'}|^2}_{\vare'}\nonumber\\
&&
\times {\rm Re}
\left[
G(\vare+i0, s\vare')G(\vare-i0, s'\vare')\right.
\nonumber\\
&&
\hspace{10mm}-\left. G(\vare+i0, s\vare')G(\vare+i0, s'\vare')
\right].
\label{eq_sigma}
\end{eqnarray}
In the second equality 
we have replaced the summation over $j$ and ${\bf k}$
with the integral in $\vare' = |\vare_{j\Vec{k}s}|$.

We first look at the Boltzmann limit by taking $\Sigma \rightarrow 0$,
which should be valid in the weak-disorder case 
satisfying $|\vare| \gg |\Sigma|$.
By assuming that $\Sigma$ is infinitesimal in (\ref{eq_self}),
we can calculate $\Sigma$ explicitly as
\begin{equation}
\Sigma(\vare + i 0) = -i {\pi \over 2} W\varepsilon_0 {\rho_0(\varepsilon) \over \rho_\infty } \equiv -i {\hbar \over 2\tau } ,
\label{eq_self_weak}
\end{equation}
where $\tau$ is the lifetime and the dimensionless parameter $W$ is defined by
\begin{equation}
W = \frac{n_i u^2}{2}\frac{\rho_\infty}{\vare_0} .
\end{equation}
For sufficiently high energy, $\hbar/2\tau\approx(\pi/2)\varepsilon_0W$
showing that $W \sim 1$ characterizes a typical disorder strength 
that smears out the fine low-energy structure due to the $k$-linear term.
This $W$ is the same as that defined for a monolayer graphene
in Ref.\ \onlinecite{Suzu} and as $A^{-1}$ in Ref.\ \onlinecite{Shon}.

For the conductivity, we take only the terms with $G^R G^A$ and $s=s'$ 
as a dominant contribution in (\ref{eq_sigma}),
and obtain a familiar form,
\begin{eqnarray}
  \sigma(\vare) \approx e^2\rho_0\tau \langle v_x^2 \rangle_\vare.
\label{eq_sigma_weak}
\end{eqnarray}
Here $\langle v_x^2 \rangle_\vare=\langle |(v_{j\Vec{k}})_{ss}|^2 \rangle_\vare$ is the Fermi-surface average 
of the diagonal matrix element of $v_x$ and use has been made of the approximation
\begin{equation}
\frac{1}{\Omega} \sum_\alpha G^R_\alpha(\vare) G^A_\alpha(\vare) 
\approx \frac{2\pi\tau}{\hbar} \rho_0.
\end{equation}

With the use of (\ref{eq_vsq_high}) the conductivity for the higher energy $|\vare| > \vare_0$ becomes
\begin{equation}
 \sigma(\vare) = \frac{e^2}{\pi^2\hbar}\frac{1}{W}\biggl(\frac{|\vare|}{\vare_0}+1\biggr).
\label{eq_sigma_weak_high}
\end{equation}
The term proportional to $|\vare|$ results from the $k$-square term of
the dispersion and is rewritten as $e^2\rho_\infty|\varepsilon|\tau/m^*$
as in the usual two-dimensional metal.  The term independent of
$|\varepsilon|$ is a correction due to the trigonal warping caused by
the $k$-linear term, which never vanishes even as
$\vare\rightarrow\infty$.  In the low-energy region $|\vare| \ll
\vare_0$, we have
\begin{equation}
\sigma(\vare) = \frac{e^2}{\pi^2\hbar}\frac{3}{4W}.
\label{eq_sigma_weak_low}
\end{equation}
The energy-independent conductivity is essentially the same as that of 
monolayer graphene.\cite{Shon}

\begin{figure}
\begin{center}
\leavevmode\includegraphics[width=80mm]{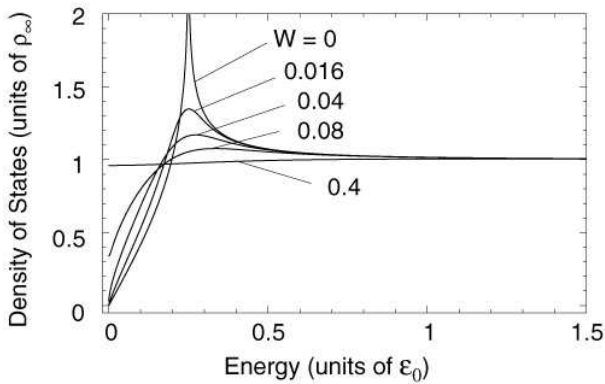}
\end{center}
\caption{Density of states per spin calculated in the SCBA.
Plots become identical for the short- and long-range scatterers.}
\label{fig_dnst}
\end{figure}

\begin{figure}
\begin{center}
\leavevmode\includegraphics[width=80mm]{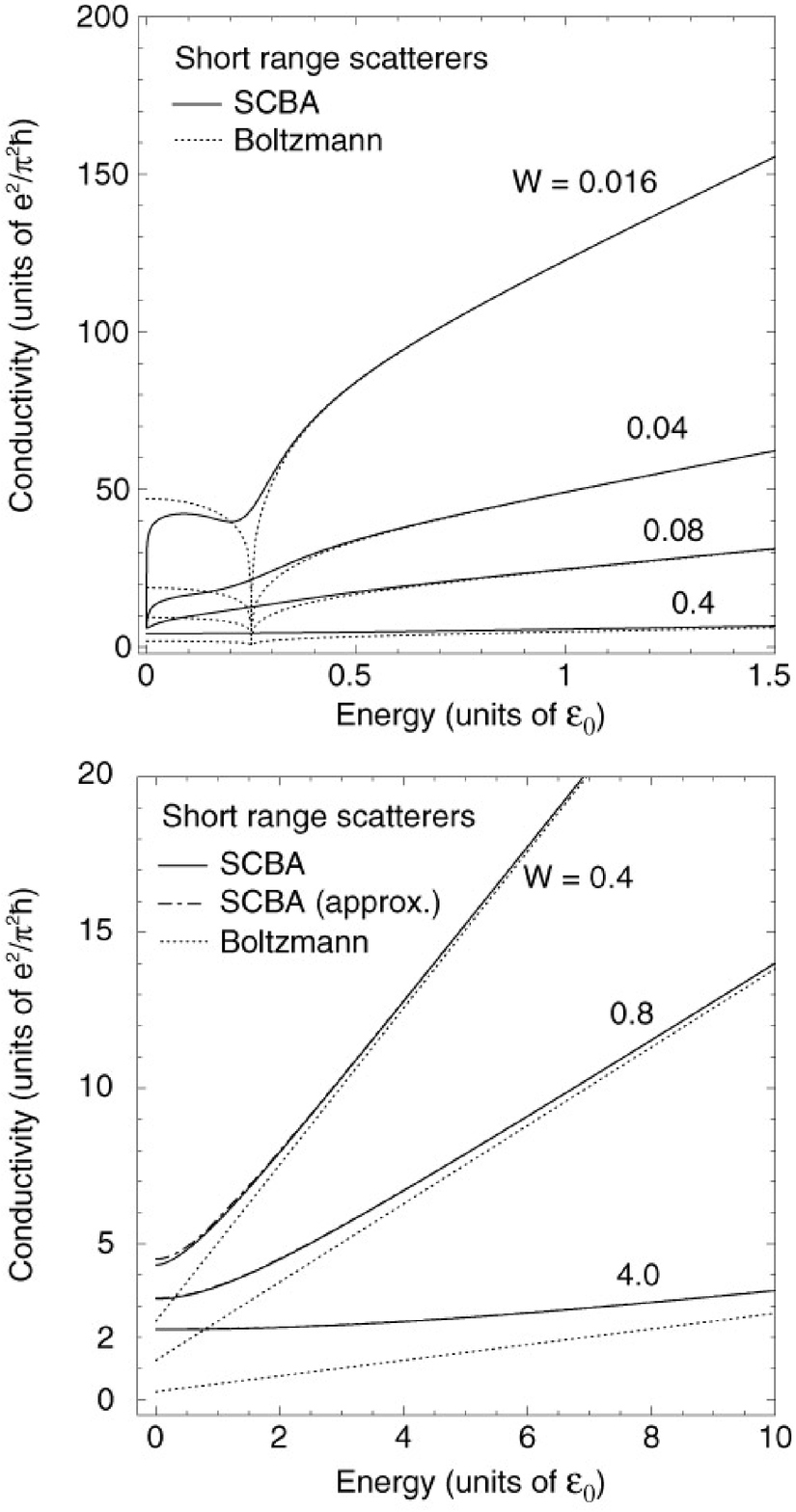}
\end{center}
\caption{Calculated SCBA conductivity (solid) and Boltzmann conductivity
(dotted) per spin in the case of short-range disorder,
with upper and lower panels showing smaller and larger $W$'s.  
For the Boltzmann conductivity in the lower panel we used the expression
(\ref{eq_sigma_weak_high}) which is valid for $\vare > \vare_0$.  In the
lower panel the approximate result given by Eq.\ (\ref{eq_sigma_scba})
is also shown.}  \label{fig_cond_short}
\end{figure}

In the case of large disorder $W > 1$, the fine structure in the density of states disappears completely (see below) and therefore the imaginary part of the self-energy becomes independent of energy and the real part becomes negligible.
Thus, we always have
\begin{equation}
 \Sigma(\vare+i0) \approx -i \Gamma ,
\label{eq_self_scba}
\end{equation}
with
\begin{equation}
 \Gamma = {\pi \over 2} W \varepsilon_0 ,
\label{eq_self_scba2}
\end{equation}
giving
\begin{equation}
\rho(\varepsilon) \approx \rho_\infty .
\end{equation}

We can calculate the conductivity (\ref{eq_sigma})
by substituting the high-energy expansions (\ref{eq_vsq_high})
and (\ref{eq_rho_high}), and obtain
\begin{equation}
  \sigma(\vare) \approx \frac{e^2}{\pi^2\hbar}
\left[ S(\vare)+1 \right]
+ \frac{e^2}{\pi^2\hbar}\frac{1}{W},
\label{eq_sigma_scba}
\end{equation}
with
\begin{equation}
S(\vare) = \left(\frac{\vare}{\Gamma} 
+ \frac{\Gamma}{\vare} \right)
\arctan \frac{\vare}{\Gamma},
\end{equation}
where terms of the order of $O(1/W^2)$ are neglected.
The first term of (\ref{eq_sigma_scba}) comes from the $k$-square term in the dispersion and the second term is a correction due to the $k$-linear term.

The conductivity goes to the Boltzmann limit (\ref{eq_sigma_weak_high}) for $\vare > \Gamma$, while for $\vare < \Gamma$ it becomes
\begin{equation}
 \sigma = \frac{e^2}{\pi^2\hbar}\left(2+\frac{1}{W}\right).
\label{eq_sigma_scba0}
\end{equation}
It is interesting that the conductivity becomes universal, i.e.,
$\sigma\rightarrow 2e^2/\pi^2\hbar$, and never vanishes in the limit of
the large disorder.  At a rough estimate, we can derive this expression
by putting the uncertainty relation $\vare \sim \Gamma$ in
(\ref{eq_sigma_weak_high}).  In Sec.\ VI, this universal conductivity
will be reconsidered in terms of Einstein's relation.


The self-consistent equation (\ref{eq_self}) can be solved easily by a
numerical iteration.
Figure \ref{fig_dnst} shows the
density of states calculated for several disorder strengths.  
We notice that the
logarithmic divergence present in $\rho_0(\varepsilon)$ at
$\varepsilon/\varepsilon_0=1/4$ is smeared out very easily, and the
structure in the vicinity of $\vare = 0$ due to the $k$-linear 
dispersion disappears around $W\sim0.1$ and is almost
unrecognizable already for $W=0.4$.

The corresponding plot for the conductivity is shown in
Fig. \ref{fig_cond_short} along with the Boltzmann limit.  In weak
disorder $W \ll 1$ we observe a sharp dip at zero energy.  This is the
analog of monolayer graphene,\cite{Shon} showing that $k$-linear
dispersion around zero energy remains intact at small disorder.  Further
discussion of the asymptotic value at $\vare =0$ will be given in Sec.\
VI.  The Boltzmann conductivity drops to zero at $|\vare| =
(1/4)\vare_0$, as the velocity vanishes at the saddle points in the
dispersion, but this singularity disappears due to the finite density of
states.  Apart from this difference the results are almost the same as
those in the Boltzmann limit.  The conductivities for large $W$'s are
plotted in a different scale in the lower panel in Fig.\
\ref{fig_cond_short}, compared with the high-energy expression of the
Boltzmann limit (\ref{eq_sigma_weak_high}) and the approximate
expression (\ref{eq_sigma_scba}).  The curves deviate from the Boltzmann
limit in the region $\vare < \Gamma$ and never fall below
$2e^2/(\pi^2\hbar)$ even in strong disorder.  The analytic expression
becomes valid already for $W\sim1/2$.

\begin{figure}
\begin{center}
\leavevmode\includegraphics[width=80mm]{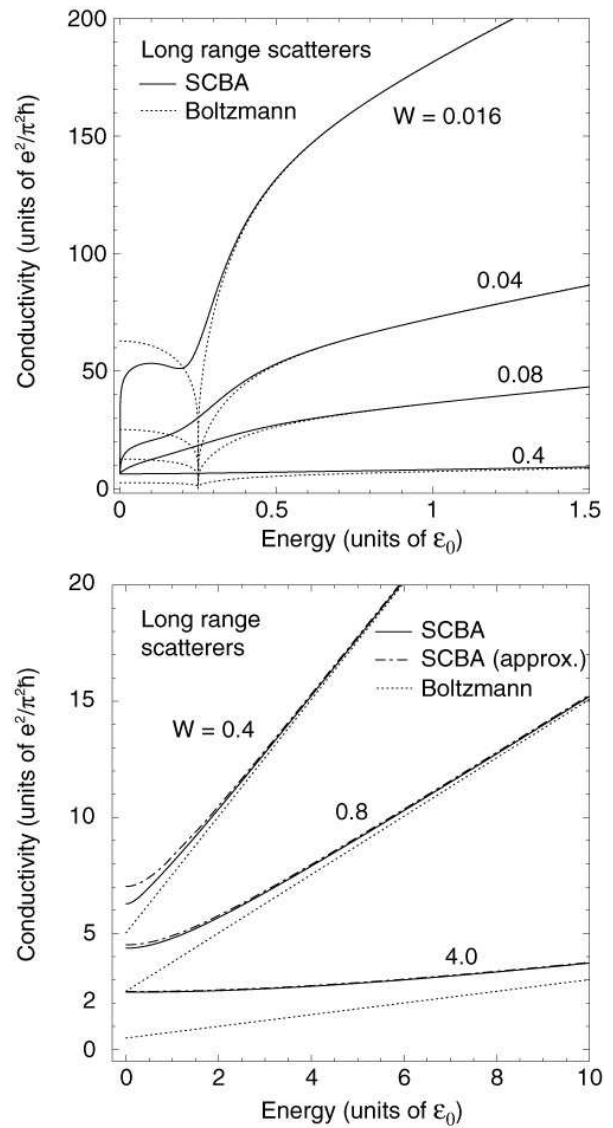}
\end{center}
\caption{Calculated SCBA conductivity (solid) and Boltzmann conductivity (dotted) per spin in the case of long-range disorder, with upper and the lower panel
showing smaller and larger $W$'s, respectively.
For the Boltzmann conductivity in the lower panel
we used the expression in (\ref{eq_sigma_weak_high_long}) which is valid
for $\vare > \vare_0$.
In the lower panel the approximate result given by Eq.\ (\ref{eq_sigma_scba_long}) is also shown.}
\label{fig_cond_long}
\end{figure}

\section{Long-range scatterers}

For the long-range disorder,
we consider scatterers with density $n_i$ and 
mean square amplitude $\av{(u_i)^2}=u^2$.
The two valleys $K$ and $K'$ are now decoupled
and the conductivity is written as the summation of their
individual contributions. 
The self-energy is given by
\begin{eqnarray}
\Sigma(\vare) = \frac{n_i u^2}{2\Omega} \sum_{\Vec{k}s} G_{j\Vec{k}s} (\vare),
\end{eqnarray}
which differs from (\ref{eq_self0}) for short-range disorder
in that the prefactor is larger by a factor of 2
and that the summation is taken only in one valley.
However those two elements cancel and the self-energy becomes 
identical with (\ref{eq_self}) in this notation.
We can omit the valley index $j$ completely.

The conductivity without the vertex correction,
denoted by $\sigma^0$, is equivalent to (\ref{eq_sigma}).
The vertex part (\ref{eq_vertex}) is given by
\begin{eqnarray}
\langle j \Vec{k}s' |  \tilde{v}_x(\vare,\vare')  | j \Vec{k}s \rangle 
&=&
\langle j \Vec{k}s' | v_x | j \Vec{k}s \rangle \nonumber\\
&& \hspace{-20mm} + {\hbar k_0 \over 4m^*}
\frac{B(\vare,\vare')}{1-\Pi(\vare,\vare')}
(s e^{-i\theta_{j\Vec{k}}} + s' e^{i\theta_{j\Vec{k}}}),
\end{eqnarray}
where $\Pi$ and $B$ are dimensionless quantities defined by 
\begin{eqnarray}
 \Pi(\vare,\vare') &=&
\frac{n_i u^2}{4\Omega} \sum_{\Vec{k}}\sum_{ss'}
G_{j\Vec{k}s}(\vare)G_{j\Vec{k}s'}(\vare') , \\
 B(\vare,\vare') &=&
\frac{n_i u^2}{4\Omega} \sum_{\Vec{k}}\sum_{ss'}
\left[1 + ss' {\rm Re} (\tilde w_{j\Vec{k}} e^{-i\theta_{j\Vec{k}}}) \right]\nonumber\\
\noalign{\vspace{-0.250cm}}
&& \hspace{20mm}\times G_{j\Vec{k}s}(\vare)G_{j\Vec{k}s'}(\vare'),
\end{eqnarray}
with $\tilde w_{j\Vec{k}} = w_{j\Vec{k}}(\hbar k_0/4m^*)^{-1}$.
The conductivity correction including the valley degeneracy
is then written as
\begin{eqnarray}
 \delta\sigma(\vare) &=& 
\frac{e^2}{\pi^2 \hbar} \frac{1}{2W} {\rm Re} 
\left(
\frac{(B^{RA})^2}{1-\Pi^{RA}}
-\frac{(B^{RR})^2}{1-\Pi^{RR}}
\right),
\label{eq_sigma_long}
\end{eqnarray}
where $B^{RA} = B(\vare+i0,\vare-i0)$ and $B^{RR} = B(\vare+i0,\vare+i0)$.

In the Boltzmann limit, only the term including 
$B^{RA}$ and $\Pi^{RA}$ is relevant.
A straightforward calculation gives
\begin{eqnarray}
\Pi^{RA} &=& \frac{1}{2} , \\
B^{RA} &=& \frac{1}{2}\av{1+{\rm Re} (\tilde w_{j\Vec{k}} e^{-i\theta_{j\Vec{k}}})}_\vare.
\end{eqnarray}
We can easily show that 
$\av{{\rm Re} (\tilde w_{j\Vec{k}}e^{-i\theta_{j\Vec{k}}})}_\vare=1$ 
to obtain 
$\delta\sigma/(e^2/\pi^2\hbar) = 1/W $ in the limit $\vare > \vare_0$.
In the case $\vare \ll \vare_0$, on the other hand, $\av{{\rm Re} (\tilde
w_{j\Vec{k}} e^{-i\theta_{j\Vec{k}}})}_\vare =0$ and therefore
$\delta\sigma/(e^2/\pi^2\hbar)=1/4W$.
The total conductivity combined with
$\sigma^0$ in (\ref{eq_sigma_weak_high}) or (\ref{eq_sigma_weak_low})
becomes
\begin{equation}
\sigma(\vare) = 
\left\{
\begin{array}{cc}
{\displaystyle \frac{e^2}{\pi^2\hbar} \frac{1}{W}
\biggl(\frac{|\vare|}{\vare_0}+2\biggr)} 
& (\vare > \vare_0) ,
\\
{\displaystyle \frac{e^2}{\pi^2\hbar} \frac{1}{W}} & (\vare \ll \vare_0).
\label{eq_sigma_weak_high_long}
\end{array}
\right.
\end{equation}

In the case of large disorder $W>1$, we can derive the analytic
expression similarly to the short-range case.  By using the self-energy
(\ref{eq_self_scba}) we can show that $B_{RA}=1$, $B_{RR}=0$, and
$\Pi_{RA}=1/2$, giving the vertex correction $\delta\sigma(\vare) =
(e^2/\pi^2\hbar)/W$ independent of energy.  The total conductivity
including $\sigma^0$ in (\ref{eq_sigma_scba}) is written as
\begin{equation}
 \sigma(\vare) = \frac{e^2}{\pi^2\hbar}[S(\vare)+1] 
+ \frac{e^2}{\pi^2\hbar}\frac{2}{W} .
\label{eq_sigma_scba_long}
\end{equation}
For $\varepsilon<\Gamma$, this reduces to
\begin{equation}
\sigma = \frac{e^2}{\pi^2\hbar}\left(2+\frac{2}{W}\right).
\label{eq_sigma_scba0_long}
\end{equation}
The conductivity is given by a universal value 
as in the short-range case in the limit of large disorder.

We show in Fig.\ \ref{fig_cond_long} the conductivity numerically
computed for the several $W$'s.  The difference from the result in the
short-range case is appreciable only in the clean limit $W\ll1$, because
the vertex correction gives only a shift of the order of
$(e^2/\hbar)/W$.  The conductivity again approaches $2e^2/(\pi^2\hbar)$
in the region $\vare < \Gamma$ in strong disorder.  The approximate
result (\ref{eq_sigma_scba0_long}) is valid for $W\agt0.8$.

\section{Discussion and Conclusion}

In Secs.\ IV and V we have seen that the conductivity in the case of
strong disorder $W>1$ deviates greatly from the Boltzmann conductivity 
and converges to the order of $2e^2/\pi^2\hbar$.
This can be understood in terms of Einstein's relation
\begin{equation}
\sigma = e^2 \rho D^* ,
\end{equation}
with the density of states $\rho$ and the diffusion constant $D^*$.
The diffusion constant is written as $D^* = \av{v_x^2}\tau$, where $\av{v_x^2}$ is the average of the squared velocity over states at the Fermi energy and $\tau$ is a relaxation time related to $\Gamma$ through the uncertainty relation $\Gamma=\hbar/2\tau$.

If we neglect the $k$-linear term in the dispersion, we have
$\av{v_x^2}=|\vare|/m^*$ and therefore $\sigma = n_s e^2\tau/m^*$ with
$n_s=\rho_\infty|\varepsilon|$.  This is nothing but the Boltzmann
conductivity for $|\varepsilon|\gg\Gamma$.  In the energy range
$|\varepsilon|<\Gamma$, however, we have $\av{v_x^2}\sim\Gamma/m^*$
because we take an average over the states $\vare_\alpha \lesssim
\Gamma$.  Thus, the diffusion constant becomes $D^*\sim\hbar/2m^*$
independent of energy using $\Gamma\tau\sim\hbar/2$.  Upon using
$\rho=\rho_\infty=m^*/\pi\hbar^2$, the conductivity becomes $\sigma \sim
e^2/\hbar$.  This conductivity is universal and independent of the band
parameters and the strength of scattering.  It is also independent of
short- or long-range disorder.

The situation becomes different in the weak-disorder limit $W \ll 1$,
where $\sigma$ at zero energy drops from the Boltzmann conductivity
almost by the factor $W$.  This behavior is essentially equivalent to
that in monolayer graphene, where the conductivity drops from the
Boltzmann to the universal value $e^2/\pi^2\hbar$ in the vicinity of the
zero energy and the near-singular drop was ascribed to a reduction of
the effective density of states contributing to the
conductivity.\cite{Shon} The universal value was shown to be unaffected
by magnetic fields,\cite{Shon} and similar near-singular behavior was
shown to be present in various transport
quantities.\cite{Suzu,Ando_and_Suzuura_2003a,Ando_2003-2004:_Quantum_anomalies}
A universal conductivity at zero energy was also reported in the square
tight-binding lattice model with one-half flux,\cite{Frad} which has
$k$-linear dispersion as well.

This zero-energy conductivity
can be explicitly estimated from (\ref{eq_sigma})
with the density of states
(\ref{eq_rho_low}) and the square velocity (\ref{eq_vsq_low}).
The integral turns out to be independent 
of the imaginary part of the self-energy 
and returns a universal value
\begin{equation}
 \sigma(0) = \frac{6e^2}{\pi^2\hbar},
\end{equation}
which is six times as large as the conductivity in the monolayer.
The extra factor comes from
the product of the density of states and the square velocity, 
$3|\vare|/(\pi\hbar^2)$, which is larger than in the monolayer
by a factor of 6 due to the existence of the satellite elliptic Fermi pockets.

Recently transport properties of bilayer graphene were studied
experimentally.\cite{Novoselov_et_al_2006a} The resistivity exhibits a
prominent peak at $\varepsilon\approx0$ and decreases rapidly with the
increase of the energy or the electron concentration.  This dependence
is explained by Eq.\ (\ref{eq_sigma_scba}) or
(\ref{eq_sigma_scba_long}) (or the lower panel of Fig.\
\ref{fig_cond_short} or \ref{fig_cond_long}) qualitatively quite well
with the disorder parameter $1\alt W\alt 2$, in the region $n_s \alt
2\times10^{12}$ cm$^{-2}$.  In particular, the observed peak resistivity
$\sim 6.5$ k$\Omega$ corresponds well to the present result for $W\sim2$
in the case of long-range disorder (dominant usually) and $W\sim1$ in
the case of short-range disorder.  However, the observed resistivity
seems to decrease much faster than that given by Eq.\
(\ref{eq_sigma_scba}) or (\ref{eq_sigma_scba_long}) with a constant $W$
for larger electron concentrations $n_s \gtrsim 2\times10^{12}$
cm$^{-2}$.  This might suggest that the effective range of the
scattering potential can be comparable to the the electron wavelength at
these electron concentrations.

We have studied the quantum transport in bilayer graphene in zero
magnetic field with the self-consistent Born approximation.  The
coexistence of the $k$-linear and $k$-square dispersion is observed as
two different behaviors in the Boltzmann conductivity, like that in the
usual two-dimensional metal and in a monolayer graphene with a massless
Dirac spectrum, which are energetically separated.  The conductivity in
the SCBA deviates from the Boltzmann limit in the case of strong
disorder $W \gtrsim 1$, and converges to $\sim 2e^2/(\pi^2\hbar)$ (per
spin).  The conductivity in the long-range scatterers exhibits a
qualitatively similar behavior to that in the short-range case, while
the vertex correction gives a positive shift of the order of $1/W$.

\section*{ACKNOWLEDGMENTS}

This work has been supported in part by the 21st Century COE Program at
Tokyo Tech \lq\lq Nanometer-Scale Quantum Physics'' and by Grants-in-Aid
for Scientific Research from the Ministry of Education, Culture, Sports,
Science and Technology, Japan.  Numerical calculations were performed in
part using the facilities of the Supercomputer Center, Institute for
Solid State Physics, University of Tokyo.

\end{document}